\documentclass[11pt,twoside]{article}
\usepackage{asp2010}

\resetcounters

\begin{document}

\title{Ages of Exoplanet Host-Stars from Asteroseismology: \\HD 17156, a Case Study}
\author{Y. Lebreton$^{1, 2}$
\affil{$^1$GEPI, Observatoire de Paris, CNRS-UMR\ 8111, 5 place Jules Janssen, 92195 Meudon, France\\
$^2$IPR, Universit\'e de Rennes 1, Campus de Beaulieu, 35042 Rennes, France}
}

\begin{abstract}
The characterization of the growing number of newly discovered exoplanets ---nature, internal structure, formation and evolution--- strongly relies on the properties of their host-star, i.e. its mass, radius and age. These latter can be inferred from stellar evolution models constrained by the observed global parameters of the host-star --- effective temperature, photospheric chemical composition, surface gravity and/or luminosity--- and by its mean density inferred from the transit analysis. Additional constraints for the models can be  provided by asteroseismic observations of the host-star. The precision and accuracy on the age, mass and radius not only depend on the quality and number of available observations of the host-star but also on our ability to model it properly. Stellar models are still based on a number of approximations, they rely on physical inputs and data that can be uncertain and do not treat correctly all the physical processes that can be at work inside a star. We focus here on the determination of the age of HD~17156, an oscillating star hosting an exoplanet. We examine the dispersion of the age values obtained by different methods ---empirical or model-dependent--- and the different sources of errors ---observational or theoretical--- that intervene in the age determination based on stellar models.
\end{abstract}

\section{Introduction}

With the advent of ultra high precision photometry as performed on board by the spatial missions CoRoT and Kepler and the HST, it is now becoming possible to detect with the same device the transit of an exoplanet in front of its host-star and the host-star oscillations. The analysis of the light curve, which carries the signature of the transiting object, provides the normalized separation $a/R_{\star}$ where $a$ is the separation between the two bodies and $R_{\star}$ the radius of the star. This quantity is a measure of the mean host-star density $\langle\rho_{\star}\rangle $, through the third Kepler law. The light curve analysis also provides the planetary radius  $R_{\rm p}$ in units of  $R_{\star}$, the impact parameter $a \cos i/R_{\star}$ and the period of the motion. Ground-based high resolution spectroscopy can provide the star radial velocity curve and in turn the planetary mass $M_{\rm p}$ in units of  $M_{\star}$. It is then necessary to estimate the mass and radius of the host-star to get the absolute mass and radius of the exoplanet and classify it as a gaseous giant body or as a telluric solid one. Furthermore to get insights on the planet internal structure, formation and evolution, we need to know not only the {\em radius} and {\em mass} of the host-star but also its {\em age} \citep{2011A&A...531A...3H}. 

Stellar ages cannot be determined by direct measurements but can only be estimated or inferred. 
As nicely reviewed by \citet{2010ARA&A..48..581S} they are many methods to estimate the age of a star according to its mass, age and configuration--- single star or star belonging to a group. Here we focus on the determination of the age of HR~17156, a pulsating star that hosts an exoplanet. This star was observed and modeled recently by \citet{2011ApJ...726....3N} and \citet{2011ApJ...726....2G} who estimated its age from comparison with stellar models. In Section \ref{obs} we present the observational data available for HR~17156. In Section \ref{emp} we present the empirical methods available to age-date the star and we revise and compare the resulting ages. In Section \ref{modeling} we discuss the ages obtained through isochrone placement in the H-R diagram. Then we proceed with a detailed modeling of the star. We concentrate on quantifying the sources of inaccuracy affecting the age inferred from stellar models which result from the choice of the input physics and parameters. In Section \ref{conclu} we synthesize the results and emphasize the progress allowed by asteroseismology.
 
\section{HD~17156: Observational Data and Constraints for the Modeling}
\label{obs}

HD~17156 (HIP~13192) is a single G5V star, located at  ${{\approx}75}$ pc according to its Hipparcos parallax \citep[$13.33{\pm}0.72$ mas,][]{2007ASSL..350.....V}. We consider the observational data of HD~17156 listed in Table\ \ref{param}. We adopt the same effective temperature $T_{\mathrm{eff}}$, surface gravity $\log g$, photospheric metallicity $\mathrm{[Fe{/}H]}$ than \citet{2011ApJ...726....2G}.  We have derived the luminosity ${L{/}L_\odot}$ from the parallax and the $V$-magnitude $V{=}8.17{\pm}0.01$, with the bolometric correction $\mathrm{BC}{=}{-}0.016$ of \cite{2003AJ....126..778V}. 

Ultra high precision photometry performed on HST has provided a light-curve carrying the signature of an exoplanet and of stellar oscillations. The transit has been analysed by \citet{2011ApJ...726....3N} who derived a measure of the mean density of the star which we express here as $M^{1/3}R^{-1}$. 
\citet{2011ApJ...726....2G} analyzed the observed oscillation spectrum, a typical p-mode solar-like one and identified 8 low-degree p-modes (see their Table\ 1). In the asymptotic approximation \citep[][]{1980ApJS...43..469T}  the frequency of a mode of radial order $n$ and angular degree $\ell{\ll}n$ writes $\nu_{n,\ell}{=}\Delta \nu\left( n{+}\frac{1}{2}\ell{+}\epsilon\right){-}\ell(\ell{+}1) D_0$. The quantity $\epsilon$ is sensitive to surface physics but weakly sensitive to the order and degree of the mode. The difference in frequency between modes of consecutive orders and same degree is approximately constant and given by $\Delta \nu{\simeq}\nu_{n+1,\ell}{-}\nu_{n,\ell}{\equiv}\Delta \nu_{\ell}$ while the difference in frequency between modes of consecutive orders and degrees differing by two units is $\delta \nu_{\ell}{\equiv}\nu_{n,\ell}{-}\nu_{n-1,\ell+2}{=}4(\ell+6)D_0$.  The differences $\Delta \nu$ and $\delta \nu$ are called the large and small separations, respectively. The large frequency separation $\Delta\nu$ is a measure of the inverse of the sound travel time across a stellar diameter and is related to the mean stellar density. The quantity $D_0$ is sensitive to the sound speed gradient in the inner regions which changes with evolution and therefore with age. The values of $\Delta\nu_0$, $D_0$ and $\epsilon$ derived from the fit of the frequencies by \citeauthor{2011ApJ...726....2G} are listed in Table\ \ref{param}.

\begin{table}[!ht]
\caption{Observational constraints for HD~17156}
\smallskip
\begin{center}
{\small
\begin{tabular}{cccccccc}
\tableline
\noalign{\smallskip}
  $T_{\mathrm{eff}}$ & $\log g$ & [Fe/H] & $L$ & ${M^{\frac{1}{3}}}{R^{-1}}$ & $\Delta\nu_0$ & $D_0$ & $\epsilon$ \\
 {[K]} & [dex] & [dex] & [$L_\odot$] & ${[M_\odot^{\frac{1}{3}}}{R_\odot^{-1}}]$ & [$\mu$Hz] & [$\mu$Hz] & [$\mu$Hz] \\
\noalign{\smallskip}
\tableline
\noalign{\smallskip}
 $6082$ & $4.31$ & $0.24$ &2.45&0.718& 83.44&0.90&1.15\\
 $\ \ 60$ & $0.04$ & $0.03$ & 0.28&0.007&\ 0.15&0.19&0.04\\
\noalign{\smallskip}
\tableline
\end{tabular}
}
\end{center}
\label{param}
\end{table}

\section{Age of HD~17156 from Empirical Methods}
\label{emp}

We discuss briefly here the empirical methods that can be used to estimate the age of HD~17156 and we report the results in Figure~\ref{ages} (right figure, second column). 

\altsubsubsection*{\bf Activity} The chromospheric activity of solar-type dwarfs is anti-correlated with their age. Empirical relations allow to rely the Ca{\textsc{II}} H \& K emission index $R^\prime_{\rm HK}=L_{\rm HK}/L_{\rm bol}$ to age \citep[see e.g.][for a recent calibration]{2008ApJ...687.1264M}. For HD~17156, \citet{2007ApJ...669.1336F} measured $\log R^\prime_{\rm HK}=-5.04$ which smallness indicates very low chromospheric activity. Using \citeauthor{2008ApJ...687.1264M}'$R^\prime_{\rm HK}$-age relation, we find an age of $7.4\pm$4.3 Gyr. Also from the  \citeauthor{2008ApJ...687.1264M} relation between the fractional X-ray emission $R^\prime_{\rm X}=L_{\rm X}/L_{\rm bol}$ and age, with $L_{\rm X}{<}28.7$ \citep{2008ApJ...687.1339K}, we estimate a lower age limit of $1.5$ Gyr for the star. 

\altsubsubsection*{\bf Gyrochronology} When they evolve, solar-type stars lose angular momentum via magnetic breaking due to their mass loss. It leads to a decrease of their rotation rate, first quantified by \citet{1972ApJ...171..565S}. 
\citet{2007ApJ...669.1167B} has proposed a new method ---gyrochronology--- to derive the age of solar-type stars from the empirical relation linking their rotation period, color and age. From \citeauthor{2007ApJ...669.1167B}'s relation as revised by \citet{2008ApJ...687.1264M} and the rotation period $P_\mathrm{rot}{=}12.8$ days measured by \citet{2007ApJ...669.1336F}, we derive an age of $1.4{\pm}0.3$ Gyr for HD~17156.

\altsubsubsection*{\bf Photospheric Lithium Abundance} At the surface of low mass stars, the lithium abundance can be depleted when the convective zone reaches the not so deep regions where Li is destroyed by nuclear reactions at $T{\approx}2.5\times10^6$ K or if mixing processes carry Li from the basis of the convective zone to the nuclear burning region. A relation between the Li-abundance, effective temperature and age is observed (but not fully understood). \citet{2009A&A...503..601B} measured the Li abundance of HD~17156 and inferred a lower age limit of $2$ Gyr on the basis of Li abundance curves as a function of $T_{\mathrm{eff}}$ published by \citet{2005A&A...442..615S}  for clusters of different ages.

\section{Age of HD~17156 from Model-dependent Methods}
\label{modeling}
The age and mass of stars can be inferred from stellar evolution models. The mass and age (i.e. the inputs of models) have to be adjusted in order to get a model that fits the observed parameters of the host-star. The more observational constraints, the better determined mass and age. We consider here two methods to age-date HD~17156: datation based on the placement of the star on pre-calculated isochrone grids ---which is a thoroughly used method--- and datation by means of a complete modeling of the star.

\subsection{Isochrone placement}
\label{iso}

This method consists in placing the star in a H--R diagram and interpreting its position by means of a grid of pre-calculated theoretical isochrones or evolutionary tracks.  In addition to the fact that the star position is defined modulo the observational error bars on luminosity, effective temperature and metallicity,  this inversion method suffers from several uncertainties. First, isochrones grids are derived from stellar models that are based on a number of assumptions/simplifications and that include uncertain or badly known input physics. Second, observations cannot be compared directly to theoretical isochrones. In particular model atmospheres, also based on assumptions and approximated physical descriptions, have to be used to derive the effective temperature, bolometric magnitude and chemical abundances of stars from spectra or colors. Furthermore different inversion techniques can be used to extract the stellar age and mass from theoretical isochrones but they have to deal with problems of degeneracy in some regions of the H--R diagram, for instance when the star is close to the zero age main sequence or at turn-off \citep[see e.g.][]{2004MNRAS.351..487P}.

\begin{figure}[!ht]
\begin{center}
\resizebox*{\hsize}{!}{\includegraphics*{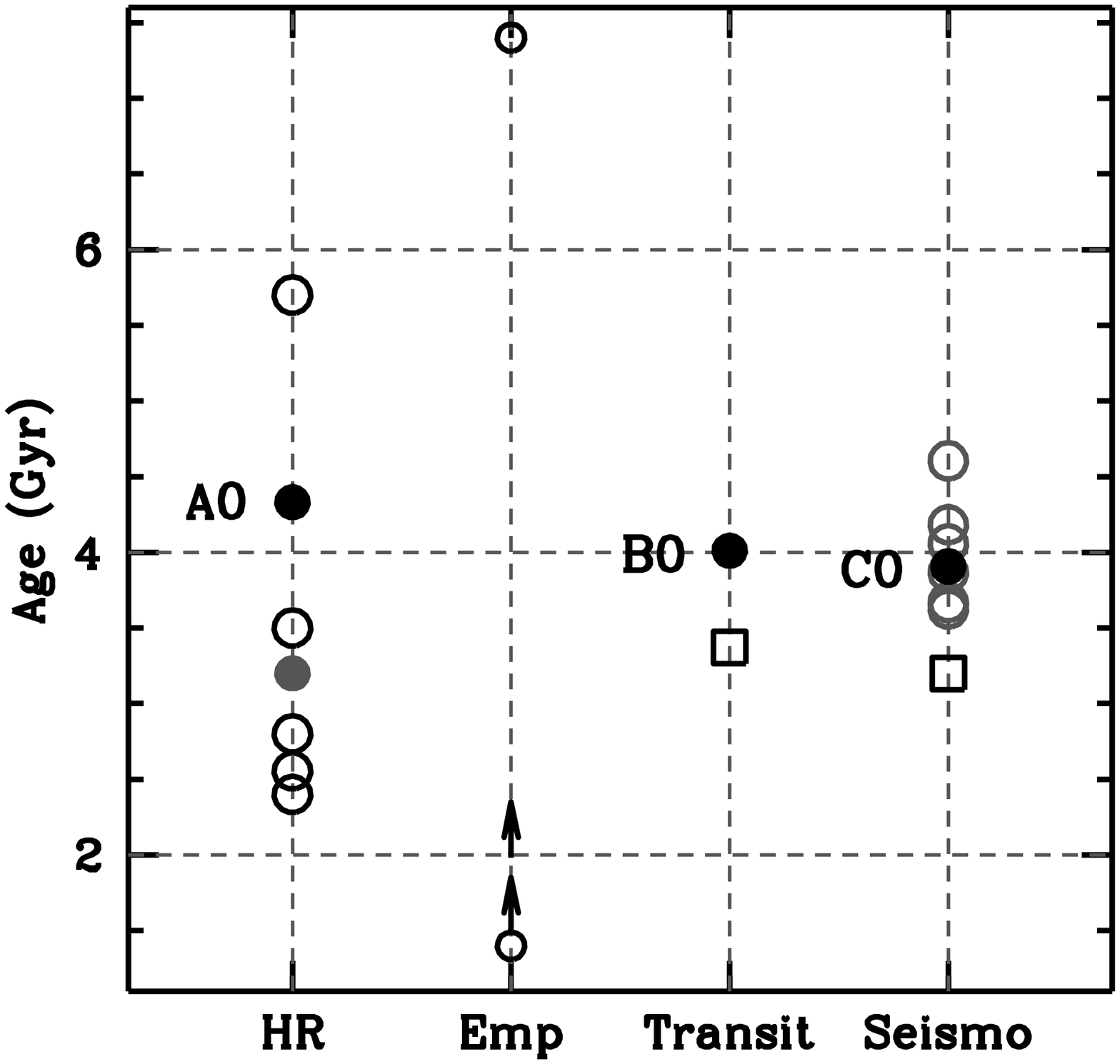}\includegraphics*{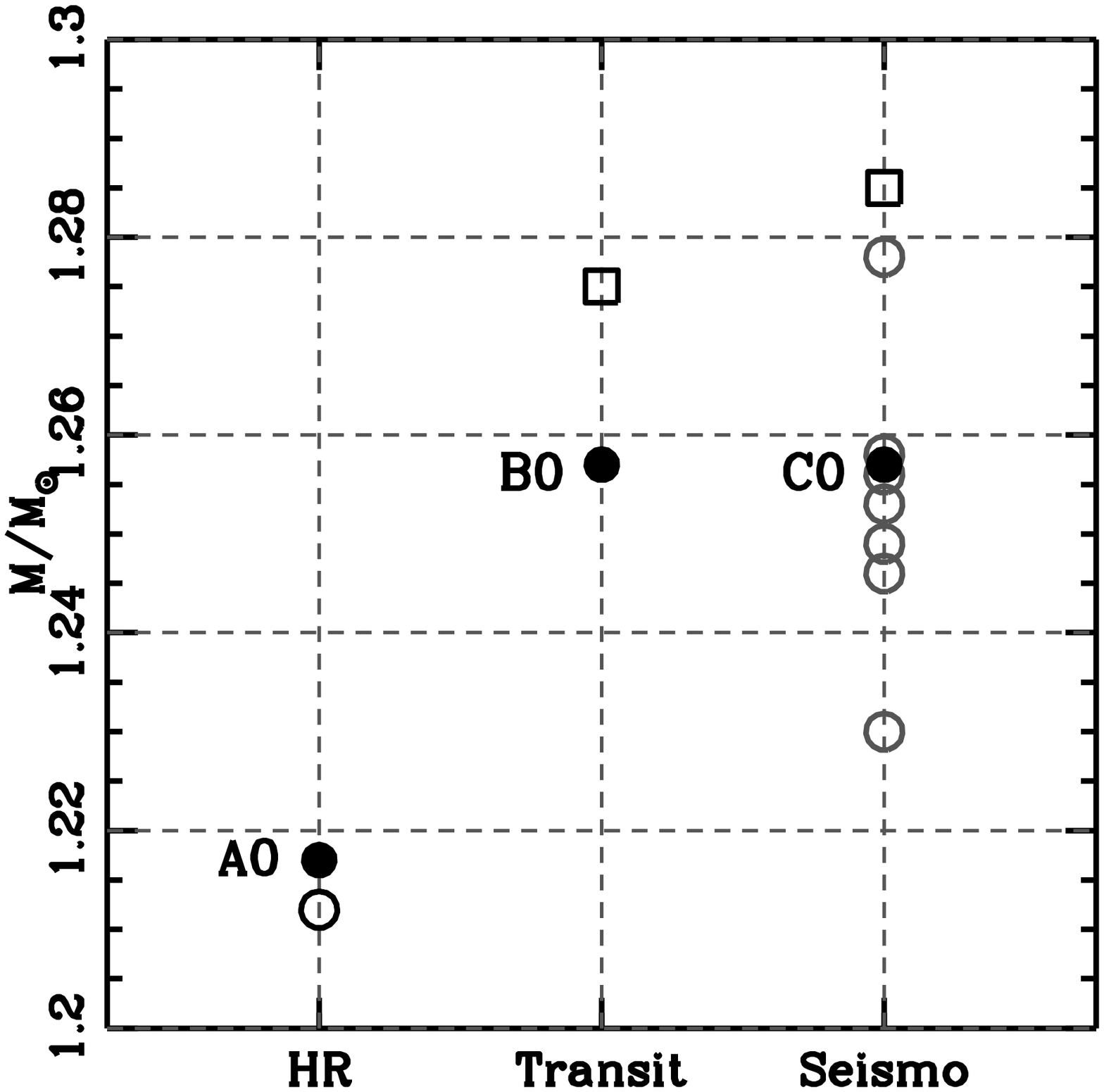}}
\caption{Age (left figure) and mass (right) estimates for HD~17156. Column ``HR'' shows estimates based on (i) inversion of isochrones by us (full grey circle) or reported from the literature (open circles) and (ii) complete modeling of the star with the H--R diagram constraints only (model A0, full black). Column ``Emp'' displays empirical ages. Column ``Transit'' shows values from complete modeling based on the H--R diagram and transit constraints (model B0, full black circle). Column ``Seismo'' are values from complete modeling based on global and seismic constraints and different sets of input physics (model C0, full black circle, models C1-7 open circles). Squares are values from \citet{2011ApJ...726....3N} and \citet{2011ApJ...726....2G}}
\label{ages}
\end{center}
\end{figure}

As for HD~17156, we derive an age of $t_\star{=}3.2{\pm}0.2$ Gyr from Padova isochrones \citep{2002A&A...391..195G} using a Bayesian inversion technique similar to the one designed by \citet{2005A&A...436..127J}. \citet{2011ApJ...726....3N} estimated an age of $3.38^{+0.20}_{-0.47}$ Gyr  by isochrone placement but incorporating the constraint on the mean-density from the transit. To summarize we show in Fig.~\ref{ages} (first column), several ages values of HD~17156 derived from isochrones inversion and picked up in the literature. Noteworthy, the ages obtained cover a large range, $2.3{-}7.2$ Gyr, due to the different isochrones grids and inversion methods used.   

\subsection{Complete Modeling}
\label{LM}

We have modeled the star with the stellar evolution code Cesam2k \citep{2008Ap&SS.316...61M} and calculated the oscillations frequencies with the {\small LOSC} pulsation code \citep{2008Ap&SS.316..149S}. We have used the Levenberg-Marquardt minimization method in the way described by \citet[][]{2005A&A...441..615M} to adjust the unknown parameters of the modeling so that the model of HD~17156 fits at best the observations, within the error bars. Our models are based on the input physics and parameters described below. If enough observational constraints are available, the free parameters can be adjusted in the modeling process, otherwise they have to be fixed ---more or less arbitrarily by the modeler (see below).

\begin{itemize}
\item {\em Microscopic Physics:} The reference models are based on the {\small OPAL05} equation of state \citep{2002ApJ...576.1064R} and on  {\small OPAL96} opacities \citep{1996ApJ...464..943I} complemented at low temperatures by {\small WICHITA} tables \citep{2005ApJ...623..585F}. We use the {\small NACRE} nuclear reaction rates \citep[][]{1999NuPhA.656....3A} except for the $^{14}N(p,\gamma)^{15}O$ reaction where we adopt the revised {\small LUNA} rate \citep{2004PhLB..591...61F}.
Models take into account microscopic diffusion of helium and heavy elements ---including gravitational settling, thermal and concentration diffusion but no radiative levitation--- following the formalism of \citet{MP93}. In alternate models we consider the {\small OPAL01} equation of state, the {\small NACRE} nuclear reaction rate for the $^{14}N(p,\gamma)^{15}O$ reaction and assume no diffusion.

\item {\em Macroscopic Physics:} We use the {\small CGM} convection theory of \citet{1996ApJ...473..550C} with a free mixing-length parameter {$\alpha_{\rm conv}$}. We consider the {\small MLT} theory \citep{1958ZA.....46..108B} as an alternative. We include convective core overshooting and assume that the overshooting zone is adiabatic and fully mixed. As a reference, we set the overshooting distance to be $\ell_{\rm ov}{=}\alpha_{\rm{ov}}\times \min(R_{\rm cc}, H_{\rm p})$ where $\alpha_{\rm ov}$, $R_{\rm cc}$ and $H_{\rm p}$ are the overshooting parameter, the radius of the convective core and the pressure scale height respectively. In alternate models, we adopt the Roxburgh \citeyearpar{1992A&A...266..291R} prescription, in which overshooting extends on a fraction of the mass of the convective core $M_{\rm cc}$, the mass of the mixed core being expressed as $M_{\rm ov}{=}\alpha_{\rm{ov}}\times{M_{\rm cc}}$. 

\item {\em Atmospheric Boundary Condition:} The reference models are based on grey model atmospheres with the classical Eddington T-$\tau$ law but we also investigate models based on the Kurucz \citeyearpar{1993yCat.6039....0K} T-$\tau$ law. 

\item {\em Solar mixture:} We adopt the revised {\small AGSS09} solar mixture \citep{2009ARA&A..47..481A} as the reference but we consider the {\small GN93} mixture  \citep{1993oee..conf...15G}  in an alternate model.
\end{itemize}

As described below we have considered different situations where $k$ unknowns of the model are adjusted to fit $k$ observational constraints. 

\paragraph{Age from Global Parameters}
In the modeling case A0 we suppose that solely the global parameters are constrained by observation ($T_{\mathrm{eff}}$, present [Fe/H], $L$) and we seek which mass, age and initial metal to hydrogen ratio $Z_0/X_0$ are required for the model to satisfy these constraints. Since this gives 3 unknowns for 3 observed parameters we have to make assumptions on the other inputs of the models, mainly the initial helium abundance $Y_0$, the mixing-length and overshooting parameters. We assume that $Y_0$ can be derived from the helium to metal enrichment ratio $(Y_0 - Y_\mathrm{P})/{Z}{=}{\Delta Y}/{\Delta Z}$ where $Y_\mathrm{P}{=}0.245 $ is the primordial He abundance \citep[e.g.][]{2007ApJ...666..636P} and ${\Delta Y}/{\Delta Z}{\approx}2$ is taken from a solar model calibration in luminosity and radius. We take $\alpha_{\rm conv}{=}0.768$ from the solar model calibration and use a moderate overshooting parameter $\alpha_{\rm ov}{=}0.15$.  The input physics are the reference ones (hereafter denoted as {\small{REF}}).

\paragraph{Age from Global Parameters and Transit}
In the modeling case B0 we add the constraint on ${M^{\frac{1}{3}}}{R^{-1}}$ ---derived from the planet transit---  to the global parameters considered in case A0 and we adjust 4 unknowns of the models (age, mass, metallicity and $\alpha_{\rm ov}$) with $\alpha_{\rm conv}$ and $Y_0$ fixed as in Case A0. The input physics are the reference ones.

\paragraph{Age from Global Parameters, Transit and Asteroseismology}
In the different cases C0--7, we add the seismic constraints  $\Delta\nu_0$, $D_0$ to the modeling and we adjust 6 model unknowns (age, mass, $Y_0$ and $Z_0/X_0$, $\alpha_{\mathrm{conv}}$, $\alpha_{\rm ov}$). We consider several alternatives for input physics: Case C0 is for reference inputs, Case C1 uses the the {\small NACRE} reaction rate for the $^{14}N(p,\gamma)^{15}O$ rate, Case C2 uses the {\small MLT} convection formalism, Case C3 is without diffusion, Case C4 uses the {\small OPAL01} equation of state,  Case C5 uses Kurucz'\ model atmospheres as boundary condition, the {\small MLT} formalism and the {\small GN93} mixture, Case C6 models overshooting as being a fraction of the convective core mass, Case C7 is based on the {\small GN93} mixture.

\begin{table}[!ht]
\caption{Parameters of HD~17156 inferred from complete modeling}
\smallskip
\begin{center}
{\small
\begin{tabular}{cccccccccc}
\tableline
\noalign{\smallskip}
  Case & Inputs &age & $M$ & $R$ & $Y_0$ &  $Z_0/X_0$ & $\alpha_{\small\mathrm{conv}}$ & $\alpha_{\small\rm ov}$ & $\chi^2$\\
&& [Gyr] &[$M_\odot$] & [$R_\odot$] & & & &  &\\
\noalign{\smallskip}
\tableline
\noalign{\smallskip}
 A0 & \small{REF}       &$4.33$ & $1.22$ & $1.41$ & $0.299$& $0.0376$ & $0.77$ & $0.15$ & $10^{-4}$\\
 B0 & \small{REF}       &$4.01$ & $1.26$ & $1.50$   & $0.298$& $0.0376$ & $0.77$ & $0.08$ & $1.3$\\
 C0 & \small{REF}       &$3.91$ & $1.26$ & $1.50$ & $0.297$& $0.0383$ & $0.72$ & $0.10$ & $1.7$\\
 C1 & \small{NACRE}  &$4.06$ & $1.25$ & $1.49$ & $0.298$& $0.0370$ & $0.73$ & $0.16$ & $1.6$\\
 C2 & \small{MLT}        &$3.67$ & $1.25$ & $1.50$ & $0.298$& $0.0374$ & $1.76$ & $0.15$ & $2.1$\\
 C3 & \small{no diffusion}     &$4.60$ & $1.25$ & $1.49$ & $0.283$& $0.0308$ & $0.69$ & $0.26$ & $3.7$\\
 C4 & \small{OPAL01} &$3.87$ & $1.26$ & $1.50$ & $0.295$& $0.0382$ & $0.78$ & $0.10$ & $3.2$\\
 C5 & \small{Kurucz}   &$4.18$ & $1.23$ & $1.49$ & $0.300$& $0.0466$ & $1.95$ & $0.13$ & $6.6$\\
 C6 & \small{$M_{\rm ov}{=}\alpha_{\rm{ov}}{\times}{M_{\rm cc}}$}   &$3.92$ & $1.26$ & $1.49      $ & $0.297$& $0.0380$ & $0.73$ & $1.18$ & $1.5$\\
 C7 & \small{GN93}     &$3.63$ & $1.28$ & $1.50$      & $0.300$& $0.0498$ & $0.73$ & $0.05$ & $1.6$\\

\noalign{\smallskip}
\tableline
\end{tabular}
}
\end{center}
\label{results}
\end{table}

\begin{figure}[!ht]
\caption{Fit of the observational constraints by the Case C0 reference model. Left: Evolutionary track in the H--R diagram with respectively the $1\sigma$ and $2\sigma$ errors bars on luminosity and temperature. The dot locates the C0-model, Centre: Similar to the previous one but for the ${M^{\frac{1}{3}}}{R^{-1}}$--$T_{\mathrm{eff}}$ plane. Right: Echelle diagram with on ordinates the frequencies of the modes and on abscissas the same frequencies modulo the large separation. Open symbols correspond to the frequencies predicted by model C0 while full symbols correspond to the 8 modes identified by \citet{2011ApJ...726....2G}. Circles, triangles and squares are for degrees $\ell{=}0, 1, 2$ respectively.}
\resizebox*{\hsize}{!}{\includegraphics*{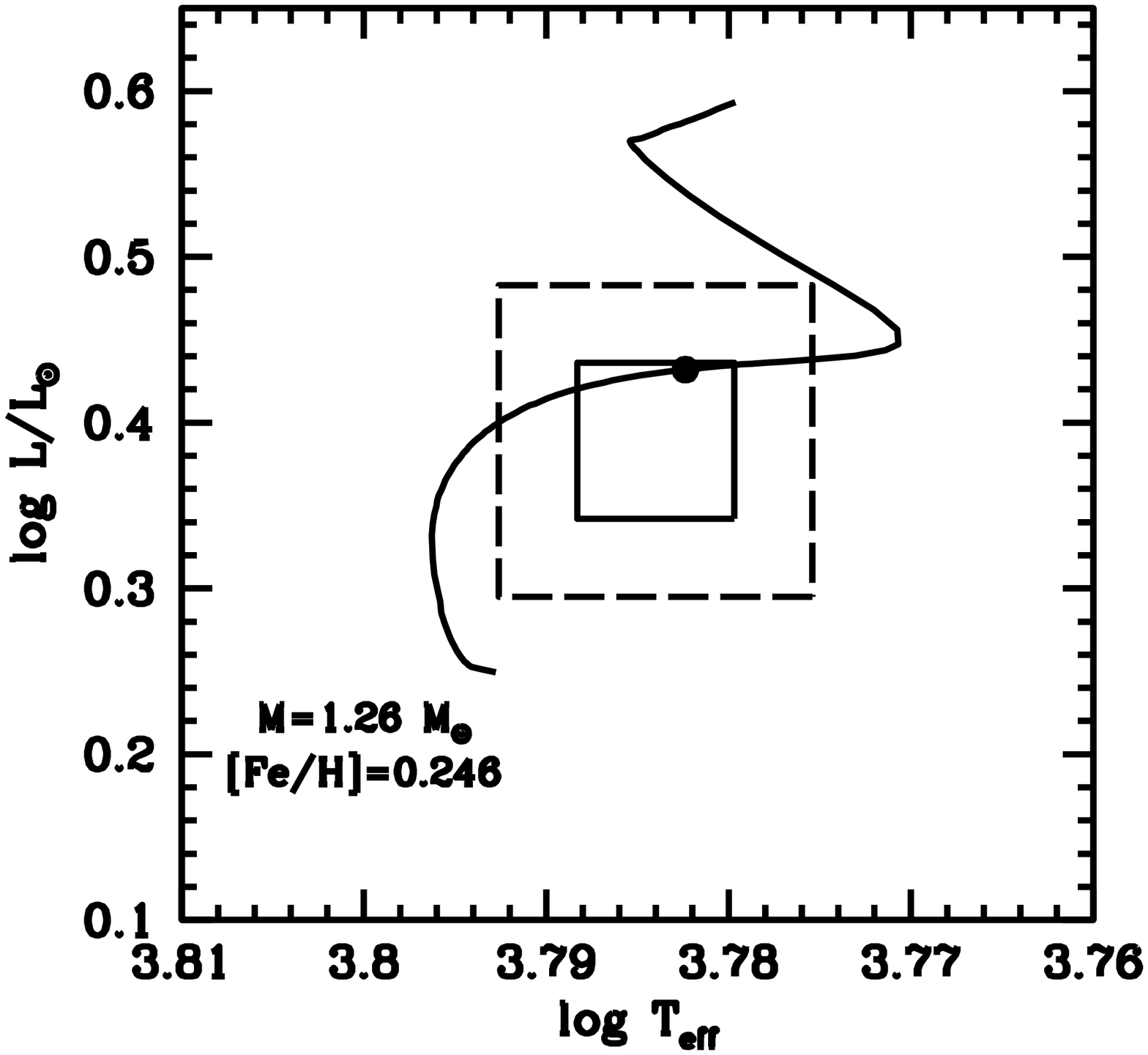}\includegraphics*{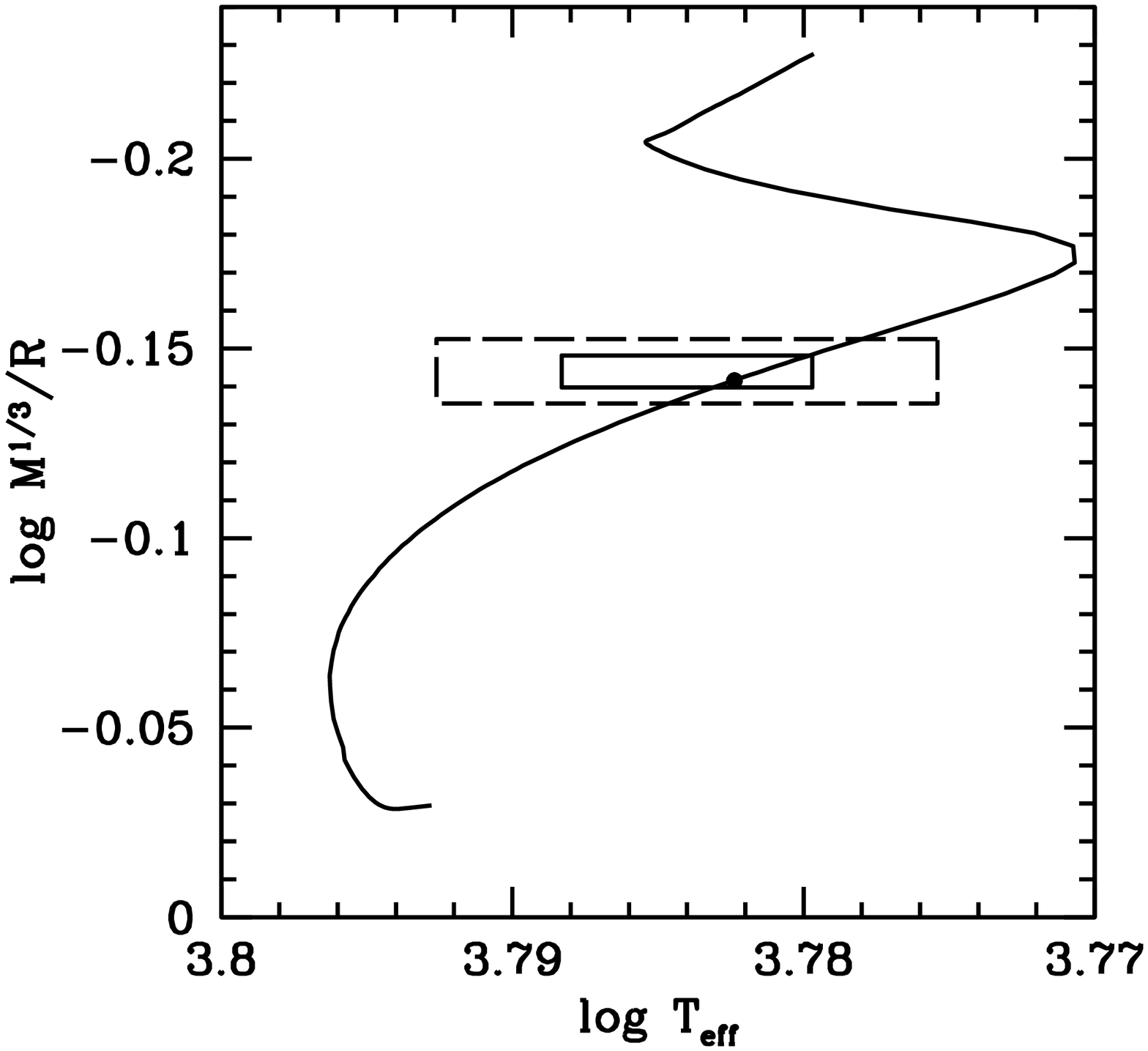}\includegraphics*{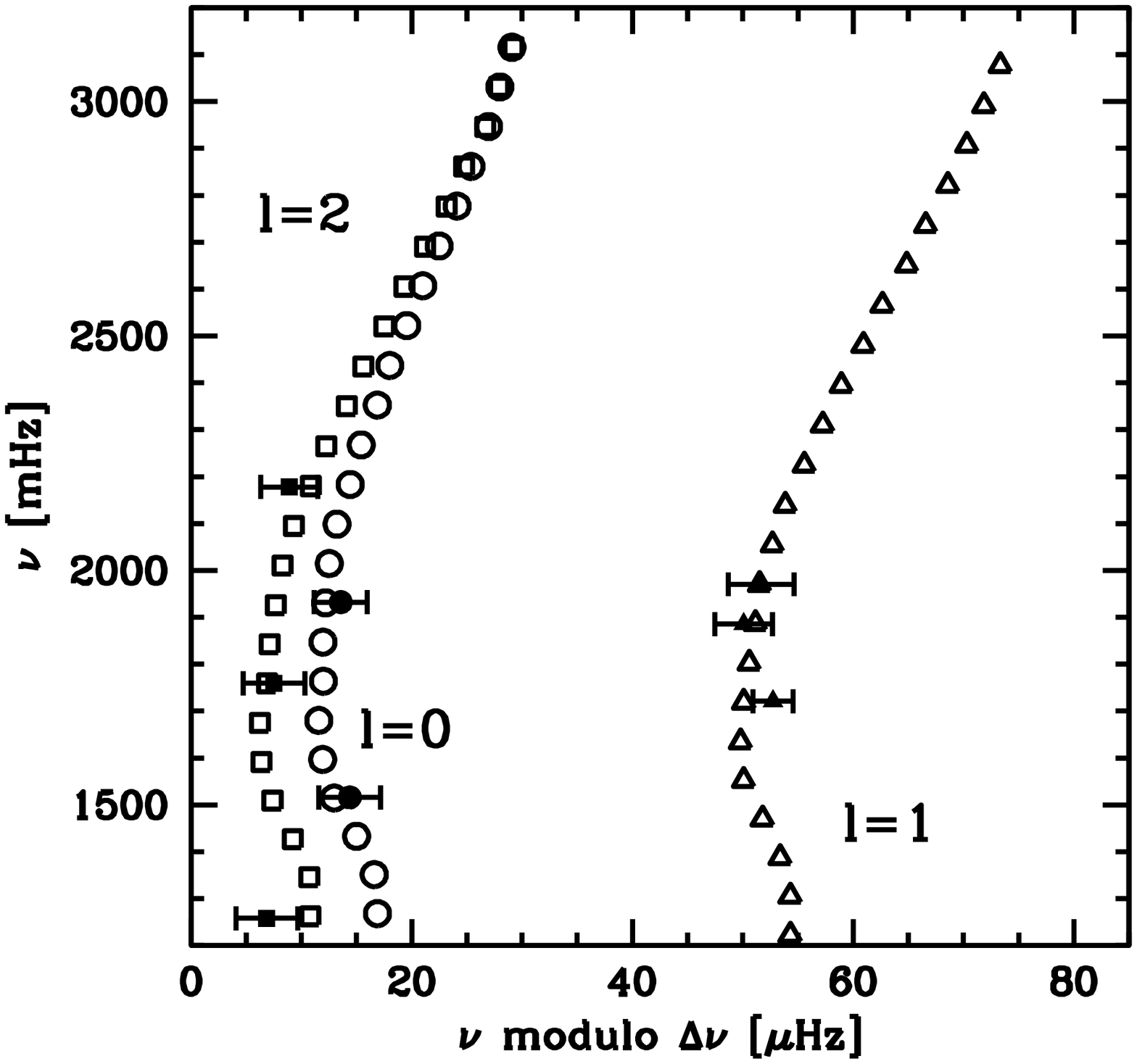}}
\label{ref-model}
\end{figure}

Table~\ref{results} lists the values of the parameters of HD~17156 estimated through complete modeling and the corresponding minimum $\chi^2$ value. The ages and masses obtained in the different cases are displayed in Fig.~\ref{ages}. In left and central Figures~\ref{ref-model} we show that the reference model C0 satisfies quite well the observed position of HD~17156  in the H--R diagram and in the ${M^{\frac{1}{3}}}{R^{-1}}$--$T_{\mathrm{eff}}$ plane. The {\'echelle--diagram} is displayed in Figure~\ref{ref-model} (right). It shows that the 8 frequencies identified by \citet{2011ApJ...726....2G} are well-fitted by the C0 model for the corresponding value of their degree.

\section{Synthesis and Conclusions}
\label{conclu}

We have investigated several methods to age-date the star HD~17156. We have found that the age estimated through empirical techniques is very ill-defined for this star which shows quite low activity. On the other hand, the age $t_\star{=}3.2{\pm}0.2$ Gyr derived from the placement of the star on a set of pre-calculated isochrones ---given its luminosity, $T_{\mathrm{eff}}$ and $[\mathrm{Fe{/}H}]$ --- has an internal error of about 10 per cent. However quite a large scatter (${\approx}3.7$ Gyr) is found when comparing the ages found in the literature because they are based on different observational parameters, inversion methods and isochrones grids.  

The complete modeling of the star presented here provides different ages depending on the number of observational constraints considered in the modeling. In Case A0, when only 3 constraints are considered (luminosity, $T_{\mathrm{eff}}$, $\mathrm{Fe{/}H}$), assumptions have to be made on the values of important model parameters (initial helium abundance, overshooting and mixing length parameters). Interestingly these latter are determined by the calibration in Cases C0-7 when seismic and transit constraints are considered. The difference in age between Case C0 ($3.91$ Gyr) and Case A0 is of $0.4$ Gyr, i.e. about 10 per cent. Furthermore we have found a scatter in age of about $25$ per cents ($1$ Gyr) when using different input physics in the models (Cases C1-7). This should be kept in mind when giving the age of stars from a given set of stellar models

By-products of the calibration are the mass $M_\star$ and radius $R_\star$ of the star. As shown in Table~\ref{results}, the differences in $M_\star$ and $R_\star$ resulting from variations of the inputs of the models are of less than $3$ and $1$ per cents respectively. The calibration also yields a rather low value of the overshooting parameter $\alpha_{\small\rm ov}{=}0.10$ which according to Roxburgh's formalism \citeyearpar{1992A&A...266..291R} corresponds to an extension of mixing on a fraction of the mass of the convective core $M_{\rm cc}$ of $18$ per cents. This will be investigated with more details in a forthcoming paper.

\acknowledgements I warmly thank Josefina Mont\'alban, Arlette Noels and Mathieu Havel for valuable advice, documents and fruitful discussions. I am very grateful to John Leibacher who presented this talk for me in Hakone.

\bibliography{ylebreton}

\end{document}